\documentclass[10pt,showpacs,amsmath,amssymb]{revtex4}

\usepackage{graphicx,color}

\def\be{\begin{equation}}
\def\ee{\end{equation}}
\def\bea{\begin{eqnarray}}
\def\eea{\end{eqnarray}}

\begin{document}

\title{Susy  $CP\sp{N-1}$ model and surfaces in ${\mathbb R}^{N^2-1}$}

\author{V. Hussin} \email{hussin@dms.umontreal.ca}
\affiliation{Centre de recherches math\'ematiques et D\'epartement de 
math\'ematiques et de statistique, Universit\'e de Montr\'eal,
C.P. 6128, succ. Centre-ville,
Montr\'eal (Qu\'ebec), H3C 3J7 Canada.} 

\author{W. J. Zakrzewski}
 \email{w.j.zakrzewski@durham.ac.uk}
\affiliation{Department of Mathematical Sciences, University of Durham, Durham DH1 3LE, United Kingdom}

\date{\today}

\begin{abstract}
\baselineskip=14pt

We describe surfaces in $R^{N^2-1}$ generated by the holomorphic solutions
of the supersymmetric  $CP\sp{N-1}$ model.
We show that these surfaces are described by the fundamental 
projector constructed out of the solutions of this model and that
in the  $CP\sp{N-1}$ case the corresponding surface 
is a sphere. Although the coordinates of the sphere are superfields
the sphere's curvature is constant. We show that for $N>2$ the corresponding
surfaces can also be constructed from the similar projector.
\end{abstract}


\maketitle

\baselineskip=14pt

\section{Introduction}
\label{intro}
\setcounter{equation}{0}
The subject of Weierstrass representations of surfaces immersed in multidimensional 
spaces was introduced few years ago by Konopelchenko et al ${}^{1,2}$. 
This has generated interest ${}^{3,4}$
in looking at the properties of these surfaces and relating
them to the solutions of the  $CP\sp{N-1}$ model.
Recently one of us (WJZ), together with Grundland ${}^{5}$
presented a general procedure for the construction of such surfaces
from the harmonic  $CP\sp{N-1}$ maps.
This approach involved writing the equation for the harmonic
map as a conservation law and then observing that in this construction
a special operator played a key role. This operator, related to the 
fundamental projector of the harmonic map was then used
in the construction of the surface.

The  $CP\sp{N-1}$ model has been supersymmetrised ${}^{6}$ thus 
giving us supersymmetric harmonic maps. The question then arises 
what surfaces these supersymmetric maps correspond to and what properties they 
have. This is the problem that is studied in this paper.

In the next section we briefly review the supersymmetric $CP\sp{N-1}$ harmonic maps (using
the formalism as given in ${}^{7}$. We then construct the operators
which are the supersymmetric generalisation of the operators of the purely 
bosonic maps. We also construct the Weierstrass surfaces and show that, like in the purely bosonic model,
 the surfaces are described by the projector of the harmonic map.
 This allows us to show that in the  $CP\sp{1}$ case, like in the corresponding
 bosonic case,  the resultant surface is the surface of a sphere.
 In the final sections of the paper we discuss its properties and present
 a short discussion of the corresponding surfaces for $N>2$.

\section{Formalism}

 \subsection{ The supersymmetric $CP\sp{N-1}$ model}

\setcounter{equation}{0}
We are interested here  in the supersymmetric (SUSY)  $CP\sp{N-1}$ model which is constructed on the two-dimensional superspace $(x,y,\theta_{1},\theta_{2})$ where the anticommuting quantities $\theta_1$ and $\theta_2$ denote two components of a majorana spinor $\bf{\theta}$ and can be thought of as being real. For our considerations, a better choice of coordinates will be the complexified superspace $(x_{+},x_{-},\theta_{+},\theta_{-})$ where
\be
x_{\pm}=x\pm i y,\quad  \theta_{\pm}=\theta_{1}\pm i \theta_{2}.
\ee
We consider in particular a complex bosonic superfield which is a $N$-column vector defined as 
\be
\Phi (x_{+},x_{-},\theta_{+},\theta_{-})= z(x_{+},x_{-})+ i \theta_{+}\  \chi_{+} (x_{+},x_{-})+ i \theta_{-}\  \chi_{-} (x_{+},x_{-})-\frac{1}{2} \theta_{+} \theta_{-} \  F(x_{+},x_{-}),
\label{phi}
\ee
where $z,\ F$ are $N$-component bosonic fields and $\chi_{+},\  \chi_{-}$ are $N$-component fermionic fields.
Since the fermionic fields $\chi_{+},\  \chi_{-}$ anticommute with each other and with $\theta_{+},\ \theta_{-}$, the hermitian conjugate of the superfield $\Phi$ is given by
\be
\Phi^\dagger (x_+,x_-,\theta_+,\theta_-)=z^\dagger (x_{+},x_{-})+ i \theta_{-}\  \chi_{+}^\dagger (x_{+},x_{-})+ i \theta_{+}\  \chi_{-}^\dagger  (x_{+},x_{-})-\frac{1}{2} \theta_+ \theta_- \ F^\dagger (x_{+},x_{-}).
\label{phidagger}
\ee
In the SUSY  $CP\sp{N-1}$ model, $\Phi $ satisfies $\Phi ^\dagger \Phi=1$. In terms of $z,\  \chi_{+},\  \chi_{-}$ and $F$, this condition writes
\begin{eqnarray} 
z^\dagger z & = &   1,
\label{condz}\\
[1ex]
\chi_{\pm}^\dagger z +z^\dagger  \chi_{\pm} & = &   0,
\label{condchi} \\ [1ex]
F^\dagger z +z^\dagger  F   & = &   2(\chi_{-}^\dagger \chi_{-} -\chi_{+}^\dagger \chi_{+} ) .
\label{condF}
\end{eqnarray} 
The usual derivatives $\partial_{\pm}=\frac12( \partial_x\pm i \partial_y)$ are generalized to superderivatives  so that we get
\be
\check\partial_\pm=-i \partial_{\theta_\pm}+\theta_\pm \partial_\pm.
\ee
They are fermionic and satisfy anticommuting properties that we have to take into account in the calculations.
For example, the following relations  will be useful later: 

\noindent
1) if $\Phi$ is a bosonic superfield, we have
\be
( \check \partial_\pm \Phi)^\dagger= \check \partial_\mp \Phi^\dagger, \ (\check \partial_+ \check \partial_- \Phi)^\dagger=\check \partial_+ \check \partial_- \Phi^\dagger.
\label{bosonic}
\ee
2) if $\Psi$ is a fermionic superfield, we have
\be
(\check \partial_\pm \Psi)^\dagger=- \check \partial_\mp \Psi^\dagger, \ (\check \partial_+ \check \partial_- \Psi)^\dagger=\check \partial_+ \check \partial_- \Psi^\dagger.
\label{fermionic}
\ee
3) In general, we have
\be
\check \partial_\pm \check \partial_\pm\,=\, - i  \partial_\pm.
\label{twice}
\ee

Let us recall that we are considering SUSY models. This means that the corresponding Lagrangian density and equations of motion must be expressed in terms of the superfields $\Phi, \  \Phi^\dagger $ and the associated supercovariant derivatives. A definition of these supercovariant derivatives has thus to be given. Let us note that they will be dependant on the superfields $\Phi$ and $\Phi^\dagger$ and will be defined as acting on bosonic as well as fermionic superfields. We get
\be
\check D_\pm \Lambda= \check \partial_{\pm}\Lambda- \Lambda A_\pm, \ A_\pm= 
\Phi^\dagger \ \check\partial_{\pm} \Phi,
\ee
where $\Lambda$ is an arbitrary homogeneous (bosonic or fermionic) superfield.
In our SUSY  $CP\sp{N-1}$  model, the quantities $A_\pm$ are scalar fermionic superfields. In particular, we have 
\be
\check D_\pm \Phi= ({\mathbb I- \mathbb P}) \check \partial_\pm \Phi,
\ee
where $\mathbb I$ is the identity operator and $\mathbb P=\Phi \Phi^\dagger$ is a projection operator.
We also have 
\be
(\check D_\pm \Phi)^\dagger=\check \partial_\mp \Phi^\dagger ({\mathbb I- \mathbb P}).
\ee
We can now write both the Lagrangian density and the equations of motion of our model as:
\be
{\cal L}= 2 ( |\check D_+ \Phi|^2-  |\check D_- \Phi|^2)
\ee
and
\be
\check D_+ \check D_- \Phi+  |\check D_- \Phi|^2 \Phi =0.
\label{equation}
\ee
Similarly to the case of the non-SUSY  $CP\sp{N-1}$ model, we can introduce the following spectral equations ($\lambda \in \mathbb R$)
\be
\check \partial_+ \Lambda=\frac{2}{1+\lambda} \ {\mathbb K}^\dagger \Lambda, \quad  \check \partial_- \Lambda=\frac{2}{1-\lambda} \ {\mathbb K} \Lambda,
\ee
where 
\be
{\mathbb K}=[\check \partial_- {\mathbb P}, {\mathbb P}], \quad  {\mathbb K}^\dagger=[\check \partial_+ {\mathbb P}, {\mathbb P}].
\label{ka}
\ee
So the equation of motion (\ref{equation}) is a compatibility condition for these spectral equations that could be written as a superconservation law :
\be
\check \partial_+ {\mathbb K}+\check \partial_- {\mathbb K}^\dagger=0.
\ee
Let us now show that ${\mathbb K}= {\mathbb M}+{\mathbb L}$ is in fact a linear combination of two distinct conserved quantities.  Indeed, since we have $\Phi^\dagger \Phi=1$, we can set 
\be
\Phi=|w|^{-1}w
\label{expphi}
\ee
and 
\be
{\mathbb P}=\Phi \Phi^\dagger= |w|^{-2}w w^\dagger.
\ee
Let us recall that we thus get
\be
tr {\mathbb P}\,=\,1,\qquad \hbox{and}\quad \det {\mathbb P}\,=\,0.
\label{nrm}
\ee

Now ${\mathbb K}=$ (\ref{ka})  takes the form
\be
{\mathbb K}=[\check \partial_- {\mathbb P}, {\mathbb P}]= |w|^{-2} (\check \partial_- w \ w^\dagger- w \  \check \partial_- w^\dagger)+ |w|^{-4} (\check \partial_- w^\dagger \ w- w^\dagger \ \check \partial_- w) w \ w^\dagger.
\ee
Setting
\be
{\mathbb M}= ({\mathbb I- \mathbb P})\ \frac{\check \partial_- w \ w^\dagger}{|w|^2}, \ 
{\mathbb L}= - \frac{w \ \check \partial_-  w^\dagger}{|w|^2}\ ({\mathbb I- \mathbb P}),
\ee
we easily get ${\mathbb K}= {\mathbb M}+{\mathbb L}$. Since we also have
\be
{\mathbb M}-{\mathbb L}=\check \partial_- {\mathbb P}, 
\ee
${\mathbb L}$ and ${\mathbb M}$ are  conserved.

Incidently the equations (\ref{equation}) when written in terms of $w$, take the form:
\be
\check \partial_+ \check \partial_- w -  \check \partial_+ w\ \frac{(w^\dagger \ \check \partial_-  w)}{|w|^2}
-   \frac{(w^\dagger \ \check \partial_+ w)}{|w|^2}  \check \partial_- w
- \frac{(w^\dagger \ \check \partial_+ \check \partial_- w)}{|w|^2}  w
+ 2 w  \frac{(w^\dagger \ \check \partial_+ w)(w^\dagger \ \check \partial_- w)}{|w|^4}=0.
\label{eqw}
\ee
\subsection{Special solutions of the equations of motion}

Let us now take $\Phi$ as in (\ref{expphi})  where we assume  that $w=w(x_+, \theta_+)$. Since, we have in this case
\be
\check \partial_- \Phi= (\check \partial_- |w|^{-1})w+|w|^{-1} \check \partial_- w= 
(\check \partial_- |w|^{-1})w
\ee
and
\be
{\mathbb P} \ \check \partial_- \Phi=\check \partial_- \Phi,
\ee
we get 
\be
\check D_- \Phi=0.
\ee

Thus we see that
\be
w=w(x_+, \theta_+)
\ee
solves the equation of motion (\ref{equation}). In analogy with the purely bosonic case
we shall call such a solution `holomorphic'. 

 In this case we also have ${\mathbb M}=0$ and 
\be
{\mathbb K}={\mathbb L}=-\check \partial_- {\mathbb P}=|w|^{-2}(-w \ \partial_- w^\dagger)+|w|^{-4}( \partial_- w^\dagger \ w)w w^\dagger.
\label{elle}
\ee

Let us now define the bosonic quantity ${\bf L}=-i \check \partial_- {\mathbb L}$ and the hermitian congugate which, from (\ref{fermionic}), is given by
${\bf L^\dagger}=i (\check \partial_- {\mathbb L})^\dagger=-i \check \partial_+ {\mathbb L}^\dagger.$
From (\ref{elle}), we  get 
\be
{\bf L}= \partial_-  {\mathbb P}, \quad {\bf L}^\dagger= \partial_ + {\mathbb P},
\ee
 so that ${\bf L}$ is conserved in the usual sense, i.e.
\be
\partial_+ {\bf L}+ \partial_- {\bf L}^\dagger=0.
\ee
Similarly as in the non-SUSY case, we can construct 
\be
{\bf X}= \int_\gamma {\bf L} \ dx_-  +\int_\gamma {\bf L}^\dagger \ dx_+,
\ee
which is independent on the contour of integration and we see that
\be 
{\bf X}\,=\, {\mathbb P}.
\ee
So, our surface is described by the projector ${\mathbb P}$.

As $tr \ {\mathbb P}=1$ not all components of ${\mathbb P}$ are independent, so if we want
to think of a vector $X$ describing our surface - we can take it as
a vector with $N\times N -1$ components constructed from the  
entries of ${\mathbb P}$.

\section{The  $CP\sp{1}$  case.} 
\label{prop}
\setcounter{equation}{0}

\subsection{Explicit form of X}

Now we look at the case of  $CP\sp{1}$ .
In this case all our original vectors have only two components. 
Thus ${\mathbb P}$ is a $2\times2$ matrix which can be written as
\be
{\mathbb P}\,=\,\left(\begin{array}{cc}{\mathbb P}_{11}&{\mathbb P}_{12}\\
{\mathbb P}_{21}&{\mathbb P}_{22}\end{array}\right)\,=\,\frac{1}{2}({\mathbb I}+X_i\sigma_i),
\label{CP1}
\ee
where 
\be
X_{1}\,=\,{\mathbb P}_{12}+{\mathbb P}_{21},\quad X_2\,=\,i({\mathbb P}_{12}-{\mathbb P}_{21}),\quad
X_3\,=\,{\mathbb P}_{11}-{\mathbb P}_{22}.
\ee

Then using (\ref{nrm}) we easily get
\be 
X_1^2\,+\,X_2^2\,+\,X_3^2\,=\,1.
\label{sphere}
\ee

This suggests that we take  for our $3$ component vector $X$ the vector whose
components are given by the quantitites given above. Given this choice we see
that our surface is the surface of a sphere or radius 1. 
 
 To get the explicit form of $X_i$ we can proceed as follows:

First,  using the overall gauge freedom, we can choose 
\be
w\,=\,\left(\begin{array}{c}1\\
W\end{array}\right) 
\label{W}
\ee
 and so we see that, effectively we are dealing 
with a bosonic superfunction $W$. Of course now ${\mathbb P}$ is given by
\be
{\mathbb P}\,=\,\frac{1}{1+\vert W\vert^2}\left(\begin{array}{cc}1&W^{\dagger}\\
 W &\vert W\vert^2\end{array}\right)
\ee 
and the components of the vector $X$ are given by  
\be
X_1\,=\, \frac{W+{W}^{\dagger}}{1+\vert W \vert^2}, \qquad 
X_2\,=\, \frac{i(W^{\dagger}-{W})}{1+\vert W \vert^2}, \qquad
X_3\,=\, \frac{1-\vert W \vert^2}{1+\vert W \vert^2}.
\label{aa}
\ee

However, these  fields are just the fields of the alternative ($S^2$) description of the  $CP\sp{1}$ model. The relation between them is given by
\be   
X_i\,=\,\Phi\sp{\dagger}\sigma_i\Phi.
\ee 

Thus the situation is the same as in the purely bosonic case.

In that case we also
knew that for holomorphic solutions of the  $CP\sp{1}$  model
the generated surface corresponded to a sphere.

Our result showing that this surface is described by the projector ${\mathbb P}$,
and then the surface vector $X$ which is constructed from this projector, in fact,
corresponds to the alternative formulation of the model, is not altered
by the supersymmetrisation of the model.

For the solutions of the equations of motion  (\ref{equation}) we have
\be
W\,=\,f\,+\,i\,\theta_{+}g,
\ee
where $f$ and $g$ are, respectively, bosonic and fermionic functions of $x_+$.

Putting all the expressions in (\ref{aa}) we see that the explicit form of the vector $X$ is given by
$$
X_1\,=\,\frac{(f+\bar{f})}{1+\vert f\vert^2}\,+\,i\theta_-\frac{\bar{g}(1-f^2)}{(1+\vert f\vert^2)^2}
\,+\,i\theta_+\frac{g(1-\bar{f}^2)}{(1+\vert f\vert^2)^2}\,+\,2\theta_+\theta_-
\frac{\bar{g}g(f+\bar{f})}{(1+\vert f\vert^2)^3},
$$
\be
X_2\,=\,i\frac{(\bar{f}-f)}{1+\vert f\vert^2}\,-\,\theta_-\frac{\bar{g}(1+f^2)}{(1+\vert f\vert^2)^2}
\,-\,\theta_+\frac{g(\bar{f}^2+1)}{(1+\vert f\vert^2)^2}\,+\,2i\theta_+\theta_-
\frac{\bar{g}g( \bar{f}-f)}{(1+\vert f\vert^2)^3},
\ee
$$
X_3\,=\,\frac{(1-\vert f\vert^2)}{1+\vert f\vert^2}\,-\,2i\theta_-\frac{\bar{g}f}{(1+\vert f\vert^2)^2}
\,-\,2i\theta_+\frac{g\bar{f}}{(1+\vert f\vert^2)^2}\,+\,2\theta_+\theta_-
\frac{\bar{g}g(1-\vert f\vert^2)}{(1+\vert f\vert^2)^3}.
$$

We note that although the components of $X$ satisfy (\ref{sphere}) they are, in fact,
superfields - {\it ie} they have fermionic parts.

\subsection{Metric}

Next we look at the metric induced on the surface and its curvature.

We introduce the metric by putting
\be
g_{ij}\,=\,\partial_iX_k\partial_jX_k,
\ee
where the sum goes over all the components of $X$.

However, it is more convenient to change variable to the holomorphic 
basis and so introduce $g_{\pm\pm}$, where the indices $+(-)$ denote the $x_+$  ($x_-$) components
of the metric. Then, as we shall see below, only the $g_{+-}=g_{-+}$ components are nonzero.

Note that as our vector $X$ is constructed from the components of ${\mathbb P}$
we have
\be
g_{\pm\pm}\,=\,tr {\partial_{\pm}{\mathbb P}\partial_{\pm}{\mathbb P}}.
\ee

Then as 
\be
\partial_+{\mathbb P}\,=\,- ({\mathbb I}-{\mathbb P})\frac{\partial_+w\,w^{\dagger}}{\vert w\vert ^2}
\ee
we see that 
\be
\partial_+{\mathbb P}\,\partial_+{\mathbb P}\,\,=\,0, 
\ee
and so we see that $g_{++}=0.$

Of course $g_{--}$ also vanishes as it is given by
$g_{--}=\bar{g}_{++}.$

However $g_{+-}$ is nonzero. To calculate it we note that its is
given by
\be
\frac{\partial_+W\,\partial_-\bar{W}}{[1+\vert W\vert^2]^2}.
\label{energ}
\ee
Note that this expression, superficially, is similar to the energy density. It would
have been it had the derivatives been $\check \partial$ and not $\partial$s.
As $W$ is a superfield $g_{+-}$ is a superfield too. So what are its components?

Clearly, the bosonic part, which comes from putting $\theta_{\pm}=0$ in (\ref{energ})
is given by
\be
\frac{\partial_+f\,\partial_-\bar{f}}{[1+\vert f\vert^2]^2}.
\label{energa}
\ee
It is the bosonic energy density {\it ie} the term that we get in a nonsupersymmetric 
version of the problem.  Calculating the other parts of $g_{+-}$ we obtain the complete result as
\be
g_{+-}=\frac{\partial_+f\,\partial_-\bar{f}}{[1+\vert f\vert^2]^2}+
 i\theta_+  \partial_+\left(\frac{g\partial_-{\bar{f}}}{[1+\vert f\vert^2]^2}\right) +
 i \theta_-  \partial_-\left(\frac{\bar{g}\partial_+{{f}}}{[1+\vert f\vert^2]^2}\right) -
 \theta_+ \theta_-  \partial_-\partial_+\left(\frac{g\bar{g}}{[1+\vert f\vert^2]^2}\right).
\label{energa}
\ee
Hence we see that the metric does have  fermionic corrections but, as they are total derivatives, they average to zero ({\it ie} vanish after integration over $x_+$ and $x_-$).

\subsection{Curvature}

 Next we calculate the curvature of our metric. As the metric has only 
 the $g_{+-}$ component  the curvature is given by
 \be
  K\,=\,-2\frac{1}{F}\,\partial_+\partial_-\,\ln F,
  \ee
  where $F=\frac{1}{2}g_{+-}$.

  To perform the calculation we note that 
  \be
  ln\left(\frac{\partial_+W\,\partial_-\bar{W}}{[1+\vert W\vert^2]^2}\right)
  \,=\, ln\left(\partial_+W)\right)\,+\, ln\left(\partial_-\bar{W}\right)\,-\, 2\,ln\left( [1+\vert W\vert^2]\right).
  \label{cu}
  \ee
  However as $W=W(x_+, \theta_+)$ only the first two terms in (\ref{cu}) vanish when one applies to them
  $\partial_+\partial_-$ and so we get
  \be
  K\,=\,-\frac{[1+\vert W\vert^2]^2}{\partial_+W\,\partial_-\bar{W}}(-2)\partial_+\partial_-
  \left(ln[1+\vert W\vert^2]\right)\,=\, 2.
  \ee
  Thus the curvature is purely bosonic and, as expected, is 2. 
 
 In a way this may be not unexpected as our surface is a surface of a sphere. However, 
 it is interesting that although the coordinates of this surface are superfields
 and the induced metric is also described by a superfield all the fermionic effects
 cancel and the curvature is just $K= 2$. Hence the fermionic modification does not
 alter the curvature of the surface.

\section{Weierstrass system for  $CP\sp{1}$ }
\label{weier}
\setcounter{equation}{0}
Let us recall the regular Weierstrass problem for the nonsupersymmetric  $CP\sp{1}$ 
system. In this case one considers two complex functions  $\psi$, $\phi$ of $x_+$ and $x_-$
which satisfy the equations
\be
\label{eq}
\partial_+ \psi\,=\,(\vert \psi\vert^2+\vert \phi\vert^2)\phi,\qquad 
\partial_- \phi\,=\,-(\vert \psi\vert^2+\vert \phi\vert^2)\psi.
\ee
Then to find a solution of these two equations one can put
\be
V\,=\,\frac{\psi}{\bar{\phi}}
\ee
and eliminate $\psi$.  Then one rewrites (\ref{eq}) as
\be
\label{eqa}
\partial_+ V\,=\,\phi^2(1+\vert V\vert^2)^2,\qquad 
\partial_- \phi^2\,=\,-2\vert \phi\vert^4 V (1+\vert V\vert^2).
\ee
 Thus 
\be 
\phi^2\,=\,\frac{\partial_+V}{(1+\vert V\vert^2)^2}
\label{phi2}
\ee
and we see that $V$ satisfies
\be
\partial_-\partial_+ V\,=\,2\,\frac{\bar{V}\,\partial_+ V\,\partial_- V}{1+\vert V\vert^2},
\ee
{\it ie} the equation of the  $CP\sp{1}$  model.

What is the supersymmetric version of this problem?  As we know, in the supersymmetric case, $V$ becomes $W$ as in (\ref{W}). Its equation of motion can be deduced easily from (\ref{W}) and (\ref{eqw}) and it is
\be
\check\partial_+\check\partial_- W\,=\, 2 \bar{W}\frac{\check\partial_+W\,\check\partial_-W}{1+
\vert W\vert^2},
\label{eqW}
\ee
Having $W$ for $V$, we take $Z^2$ as the supersymmetric analogue of $\phi^2$ defined in (\ref{phi2}). We require that $W$ and $Z^2$ satisfy 
\be
\check \partial_+ W\,=\,(1+\vert W\vert^2)^2\, Z^2,\qquad 
\check \partial_+ Z^2\,=\,-2W\,Z^2\,\bar{Z}^2\,(1+\vert W\vert^2).
\label{wei}
\ee
Note that $W$ is bosonic while $Z^2$ is fermionic.
We have
\be
Z^2\,=\,\frac{\check\partial_+ W}{(1+\vert W\vert ^2)^2}
\ee
and, as is easy to check,  $W$ solves the equation (\ref{eqW}).

Can one take the nonsupersymmetric limit of this problem?
This is difficult as $Z^2$ is fermionic.
However, we can put
\be
\phi^2\,=\,\check\partial_+ Z^2.
\ee
Then,  as 
\be 
\check \partial_+\bar{W}\,=\,0
\ee
we see that
\be
\check \partial_+Z^2\,=\,\frac{\check \partial_+\check \partial_+ W}{[1+\vert W\vert^2]^2}.
\ee 

Note that due to (\ref{twice}) we see that (up to an over factor $-i$) this is
the correct expression for $\phi^2$ after we have set all $\theta_{\pm}=0$.

\section{Generalisation to  $CP\sp{N-1}$ }
\label{gener}
\setcounter{equation}{0}

Some of our results generalise easily to the  $CP\sp{N-1}$  case.
This is the case in particular with the projector ${\mathbb P}$ which gives
us a surface in $R^{N^2-1}$ for the  $CP\sp{N-1}$  model.

So our surface is defined in terms of ${\mathbb P}$. How should we then define our 
vector $X$? A little thought shows that, like in the nonsupersymmetric case, we should
take $X$ in such a form that an analogue of (\ref{CP1}) holds, {\it ie} $
\partial_-X_i\partial_+X_i$ is proportional to $tr \partial_-{\mathbb P}\partial_+{\mathbb P}$
as then $g_{++}=g_{--}=0$.

This requires that we take off-diagonal entries of matrix ${\mathbb P}$, say ${\mathbb P}_{ij}$
and form from them components ${\mathbb P}_{ij}+{\mathbb P}_{ji}$ and $i({\mathbb P}_{ij}-{\mathbb P}_{ji})$. For the diagonal entries we have some choice. We want the $N-1$ vector components
$X_i$ to be such that
\be 
\sum_{i=1}^{N-1}\partial_+X_i\partial_-X_i\,=\, 2\, \sum_{i=0}^N\partial_+{\mathbb P}_{ii}\partial_-{\mathbb P}_{ii}.
\ee
In the ${\mathbb C}P^1$ case this tells us that we should take, as shown in (\ref{CP1}), $X_3={\mathbb P}_{11}
-{\mathbb P}_{22}$. For larger $N$ we have more choices; thus for  $CP\sp{2}$  we can 
take (this choice is based on Gell Mann's $SU(3)$ $\lambda$ matrices)
\be 
X_1\,=\,{\mathbb P}_{11}\,-\,{\mathbb P}_{22}, \qquad  X_2\,=\,\sqrt{3}({\mathbb P}_{11}\,+\,{\mathbb P}_{22}).
\ee
or we could make another choice. In general, for  $CP\sp{2}$  we could take
\be
{\mathbb P}_{11}\,=\, \frac{1}{3}\,+\,a X_1\,+\, bX_2,\qquad {\mathbb P}_{22}\,=\,
 \frac{1}{3}\,+\,cX_1\,+\, dX_2.
\ee

Then we choose $a$, $b$, $c$ and $d$ so that
\be 
\partial_+X_{1}\partial_-X_{1}+\partial_+X_{2}\partial_-X_{2}
\ee
give the same expression as
\be
\partial_+{\mathbb P}_{11}\partial_-{\mathbb P}_{11}+\partial_+{\mathbb P}_{22}\partial_-{\mathbb P}_{22}+
\partial_+{\mathbb P}_{33}\partial_-{\mathbb P}_{33}
\ee
in which we can eliminate ${\mathbb P}_{33}$ by ${\mathbb P}_{33}=1-{\mathbb P}_{11}-{\mathbb P}_{22}.$

This guarantees that only $g_{+-}$ is nonzero.
A simple calculation shows that we have a one-parameter family of solutions
\be
a\,=\,\frac{2}{\sqrt{3}}\cos\alpha,\quad b\,=\,\frac{2}{\sqrt{3}}\sin\alpha,
\ee

$$
c\,=\,\mp\sin\alpha\,-\frac{1}{\sqrt{3}}\cos\alpha\,,\quad d\,=\,-\frac{1}{\sqrt{3}}\sin\alpha\,\pm \cos\alpha.
$$

For $N>2$ the solutions are even more nonunique.
 
Note also that with all these choices  we always have 
\be
g_{+-}\,=\,tr(\partial_+{\mathbb P}\,\partial_-{\mathbb P}).
\label{mett}
\ee

Moreover, the other components of the metric vanish.
Thus the metric has a nontrivial dependence on the fermionic degrees of freedom.
A simple calculation shows that we can rewrite (\ref{mett})
as 
 \be
g_{+-}\,=\,(\partial_+\Phi^{\dagger}\,({\mathbb I}-{\mathbb P})\partial_-\Phi)+(\partial_-\Phi^{\dagger}\,({\mathbb I}-{\mathbb P})\partial_+\Phi).
\label{metta}
\ee
This is closely related to the energy density of the original map - in the nonsupersymmetric case
it is proportional to this density; this is not the case here
as (\ref{mett}) involves $\partial$ derivatives and not $\check\partial$!

It is easy to see that the fermionic contributions to both the metric and the curvature do not cancel.
We have looked at these corrections in the  $CP\sp{2}$ case. Then the vector
$w$ has three components which can be taken in the form
\be
w\,=\left(\begin{array}{c}1\\
W_1\\
W_2\end{array}\right). 
\ee  
The detailed calculations then show that $g_{+-}$ is again given by the same expression 
as the energy density of the nonsupersymmetric model with, however, superfields
in place of bosonic fields. Thus
\be
g_{+-}\,=\,
\frac{\vert\partial_+W_1\vert^2+\vert \partial_+W_2\vert^2+\vert W_2\partial_+W_1-W_1\partial_+W_2
\vert^2
}{[1+\vert W_1\vert^2+\vert W_2\vert^2]^2}.
\label{energtwo}
\ee

We can now expand this expression in powers of $\theta$. However, it is easy to check
that as, say, the $\theta_+$ corrections involve expressions that are not
total derivatives. The same is true for the calculation of the curvature. In the  $CP\sp{1}$  case 
we had the nice factorisation of the terms in $g_{+-}$ leading to the fact that
the derrivative terms did not contribute to $\partial_+\partial_-\,ln(g_{+-}).$
This was essential for the cancellation of various factors leading to $K=2$.
This time the numerator in (\ref{energtwo}) contains 3 terms and it does
contribute to $\partial_+\partial_-\,ln(g_{+-}).$ In consequence $K$ is not very simple
and the fermionic contributions to it do not cancel. We have checked this explicitly
but as the obtained expression is quite complicated we do not present it here.
Hence, the simple results of the  $CP\sp{1}$ case do not hold any more; both
the metric and its curvature are given by full superfields.

\section{Conclusions}
\label{concl}
\setcounter{equation}{0}

In this paper we have discussed the supersymmetrisation of the Weierstrass
problem and extended to the supersymmetric case the work of Grundland et al ${}^{5}$.
Our results have shown that with small modifications the extension has not lead
to results which are significantly different from the purely bosonic case.
In the  $CP\sp{1}$  case we have again obtained a sphere.  
Its coordinates are given by real bosonic superfields and, as such, this sphere,
is deformed but its fermionic structure. However, these fermionic fields
do not play a role in the description of some of its properties; {\it eg} in the calculation
of the curvature all the fermionic contributions cancel and, as in purely bosonic case,
we get $K=2$. They do play a role in the metric - but as they are given by total
derivatives, they cancel when we integrate over $x_+$ and $x_-$.

When taking larger $N$
we have found that, for the holomorphic  $CP\sp{N-1}$  fields,
the projector ${\mathbb P}$ still describes
the surfaces in $R^{N^2-1}$. This time, however, the curvature is not constant
and, furthermore, it contains fermionic corrections.

The more general solutions of the supersymmetric  $CP\sp{N-1}$  model, for $N>2$,
are given by fields which are neither holomorphic nor antiholomorphic. Their description is
 somewhat complicated due to the constraints of the model. The
 corresponding surfaces are expected to be more complicated. They have not been 
 studied yet due to these constraint problems which still have to be resolved.
 This work is currently under consideration.

\begin{acknowledgments}
The work reported in this paper was started when V.~Hussin visited 
the University of Durham in Michaelmas term 2005.
She would like to thank the University of Durham for the award of a
Grey College Fellowship and the Departement of Mathematical Sciences for its
hospitality. 

The research of V.~Hussin is partially supported by research grants  from NSERC of Canada. 

This paper was finished when W. J.~Zakrzewski visited the University of Montreal
in March 2006. He wishes to thank the University of Montreal for hospitality and
 the LMS for its travel grant. 
\end{acknowledgments}

\vskip0.8cm

%%%%%%%%%%%%%%%%%%%%%%%%%%%%%%%%%%%%%%%%%%%%%%%%%%
%%%%%%%%%%%  References  %%%%%%%%%%%%%%%%%%%%%%%%%%%%%%
%%%%%%%%%%%%%%%%%%%%%%%%%%%%%%%%%%%%%%%%%%%%%%%%%%
{\footnotesize
\baselineskip=14pt

%\bibitem{GZ2}
%A.M. Grundland and W.J. Zakrzewski, Geometric Aspects of $CP\sp{N-1}$ Harmonic Maps, {\it J. Math. %Phys.} {\bf 44}, 328-337 (2003).

\noindent
${}^{1}$
B. Konopelchenko and I. Taimanov, Constant mean curvature surfaces via
an integrable dynamical system, {\it J. Phys.} {\bf A 29}, 1261-1265 (1996).

\noindent
${}^{2}$
R. Carroll and B. Konopelchenko, Generalised Weierstrass-Enneper inducing
conformal immersions and gravity, {\it Int. J. Mod. Phys.} {\bf A 11}, (7), 
1183-1216 (1996).

\noindent
${}^{3}$
B. Konopelchenko and G. Landolfi, Generalised Weierstrass representation
for surfaces in multi-dimensional Riemanian spaces, {\it Stud. Appl. Maths.}
{\bf 104}, 129-169 (1999) and references therein.

\noindent
${}^{4}$
 P. Bracken and A.M. Grundland, Symmetry properties and explicit solutions
of the generalised Weierstrass system, {\it J. Math. Phys.} {\bf 42}, 1250-1282
(2001) and references therein.

\noindent
${}^{5}$
A.M. Grundland, and W.J. Zakrzewski, 
$CP^{N-1}$ harmonic maps and the Weierstrass
problem, {\it J. Math. Phys.} {\bf 44}, 3370-3382 (2003).

\noindent
${}^{6}$ 
A. D'Adda, M. Luscher and P. Di Vecchia, Confinement and  chiral symmetry breaking
in $CP^{N-1}$ models with quarks, {\it Nucl. Phys.}, {\bf B 152}, 125-144, (1979).

\noindent
${}^{7}$
 see {\it eg} W.J. Zakrzewski, {\em Low Dimensional Sigma Models\/} (Hilger, Bristol, 1989).

\end{document}